%% file: ms.tex
\documentclass[preprint]{aastex}
\usepackage[dvips]{color}
\newcommand{\re}{\textcolor[rgb]{0,0,0}}
\newcommand{\ree}{\textcolor[rgb]{0,0,0}}
\newcommand{\bl}{\textcolor[rgb]{0,0,0}}
\newcommand{\bu}{\textcolor[rgb]{0,0,0}}
\newcommand{\ma}{\textcolor[rgb]{0.0,0,0.0}}

\def\gtsima{$\;\buildrel > \over \sim \;$}
\def\simgt{\lower.5ex \hbox{\gtsima}}
\def\ltsima{$\;\buildrel < \over \sim \;$}
\def\simlt{\lower.5ex \hbox{\ltsima}}
\usepackage{lineno}

\begin{document}

\title{SOFIA/EXES observations of warm H$_2$ at high spectral resolution: II.\ IC443C, NGC2071, and 3C391 }

\author{David A. Neufeld\altaffilmark{1}, Curtis DeWitt\altaffilmark{2}, Pierre Lesaffre\altaffilmark{3,4}, Sylvie Cabrit\altaffilmark{4}, Antoine Gusdorf\altaffilmark{3,4},
Le Ngoc Tram\altaffilmark{5}, and William T.\ Reach\altaffilmark{2}}

\altaffiltext{1}{Department of Physics \& Astronomy, Johns Hopkins Univ., Baltimore, MD 21218, USA}
\altaffiltext{2}{Space Science Institute, 4765 Walnut Street, Suite 205, Boulder, CO 80301, USA}
\altaffiltext{3}{\bu{Laboratoire de Physique de l'\'Ecole Normale Sup\'erieure, ENS, Universit\'e PSL, CNRS, Sorbonne Universit\'e, Universit\'e Paris Cit\'e, F-75005, Paris, France}}
\altaffiltext{4}{Observatoire de Paris, PSL University, Sorbonne Universit\'e, LERMA, 75014, Paris, France}
\altaffiltext{5}{Max-Planck-Institut f\"ur Radioastronomie, Auf dem H\"ugel 69, D-53121, Bonn, Germany}

\begin{abstract}

Using the EXES instrument on SOFIA, we have obtained velocity-resolved
spectra of several pure rotational lines of H$_2$ toward shocked molecular gas within
three Galactic
sources: the supernova remnant (SNR) IC443 (Clump C), a protostellar outflow in the intermediate-mass star-forming region NGC 2071, and the SNR 3C391.  
These observations had the goal of searching for expected velocity shifts between 
ortho- and para-H$_2$ transitions emitted by C-type shocks.
In contrast in our previous similar study of HH7, the result of our search was negative: no velocity shifts were reliably detected.  \re{Several possible explanations for the absence of such shifts are discussed: these include a preshock ortho-to-para ratio that is already close to the high-temperature equilibrium value of 3 (in the case of IC443C), the more complex shock structures evident in all these sources, and the larger projected aperture sizes relative to those in the observations of HH7.}

\end{abstract}


\section{Introduction}

Previous observations of shocked interstellar gas, performed with SOFIA/EXES toward the classic bow shock in HH7 (Neufeld et al.\ 2019, hereafter N19), have provided key constraints on shock physics through velocity-resolved measurements of H$_2$ rotational line profiles.  These revealed, for the first time, clear velocity shifts, 
in the range 3 -- 4~km/s, between emission lines of ortho-H$_2$ (the S(5) and S(7) transitions) and those of para-H$_2$ (the S(4) and S(6) transitions).  As discussed by N19, 
to which the reader is referred for a more extensive discussion, these shifts are naturally explained as resulting from 
the relative slow conversion of para-H$_2$ to ortho-H$_2$ within the region of elevated gas  temperature within an interstellar shock; as such, they provide perhaps the 
most direct evidence for 'C-type' shocks in which the flow velocity varies continuously within the warm shocked gas \bl{thanks to ion-neutral collisions in the magnetized medium
(Draine 1980; Chernoff, Hollenbach \& McKee 1982; Flower, Pineau des For\^ets \& Hartquist 1985).}  Within the velocity range covered by the emission line, the ortho-to-para ratio (OPR, \re{the determination of which is discussed in Section 4 below}) was found to vary smoothly between $\sim$ 0.5 and 2.0 \bl{across the line profile}, in good agreement with the predictions of C-type shock models in which cold ambient gas, with
a low initial OPR, is accelerated and heated; this is accompanied by partial para-H$_2$ to ortho-H$_2$ conversion towards the equilibrium OPR of 3 expected in warm gas.

Motivated by this earlier result, we have observed the S(4) -- S(7) rotational lines of H$_2$ with SOFIA/EXES toward three additional sources where shock-heated gas is present.  These were chosen to reflect a diversity of interstellar shock types, all rather 
different from that in HH7, and comprise the supernova remnants (SNR) IC443 (Clump C) and 3C391, along with a protostellar outflow in the intermediate-mass star-forming region NGC 2071.  \ree{All three sources had been observed with the Spitzer Space Telescope, which obtained
spectral line maps for the S(0) - S(7) H$_2$ rotational lines (Neufeld et al.\
2007, 2009).  Full descriptions of these sources were given in those earlier
publications and will not be repeated here.   At the spectral resolving power of Spitzer Infrared Spectrometer (IRS), $\lambda/ \Delta \lambda = 60$ for the S(3) - S(7) lines and $\lambda/ \Delta \lambda = 600$ for the S(0) - S(2) lines, the H$_2$ rotational line profiles could not be measured; moreover, because of the limited spectral resolution,
the S(6) line could not be observed reliably toward 3C391 because of a strong
PAH emission feature around 6.2$\mu$m.}

\ree{As in several other shocked sources observed with Spitzer, the H$_2$ rotational diagrams observed toward all three sources could be fit successfully with a two component model that invoked a warm (250 - 350 K) gas component
responsible for the lower-lying transitions and a hot (1000 - 1100 K) component
responsible for the higher-lying transitions.   Such models
yield best estimates of 0.52 and 0.65 for the OPR in the warm components in  NGC 2071 and 3C391, respectively, far lower than the value $\sim 3$ expected in equilibrium at the
measured gas temperature; the low H$_2$ OPRs measured for the warm gas in these sources are clearly a relic of an earlier epoch when the gas had reached equilibrium at a lower temperature.  In the absence of flux measurement for the S(6) line, an OPR could not be obtained for the hot gas component in 3C391, while that observed for the hot component
in NGC 2071 was consistent with 3. For IC443C, by contrast, the  S(0) - S(7) line ratios were entirely consistent with an equilibrium OPR of 3 for {\it both} the warm and hot components.  These Spitzer observations suggested that NGC 2071 and 3C391 were good candidates for the detection of ortho-para line shifts like those measured toward HH7.
IC443C, on the other hand, was selected as a control in which no such shifts were 
expected.}

In Section 2 below, we describe the observations and data reduction.  The resultant spectra are presented in
Section 3.  They are discussed in Section 4, where the gas temperature and OPR is determined as a function of velocity along the line-of-sight.   

\section{Observations and data reduction}

We observed the S(4), S(5), S(6), and S(7) pure rotational transitions of H$_2$ toward a single position in each of IC443C, 3C391, and NGC 2071.  In each case, \ree{to
maximize the observed signal}, the EXES slit was centered at the peak of the H$_2$ S(5) emission, as revealed by the previous line mapping observations with {\it Spitzer}/IRS \ree{and without regard to the OPR ratio}.  
Although the IRS spectrometer lacked the spectral resolution to resolve the H$_2$ line profiles, it allowed maps of size 
$\sim 1^\prime \times 1^\prime$ to be constructed by scanning the spectrometer slit perpendicular to its length.  In Figures 1 -- 3, we show for each source the location and position angle of the EXES slit on the Spitzer maps of the H$_2$ S(5) intensity.  The
observing parameters
are given in Table 1, along with the wavelength and upper state energy for each line (Roueff et al. 2019).

\begin{figure}
\includegraphics[scale=0.7,angle=0]{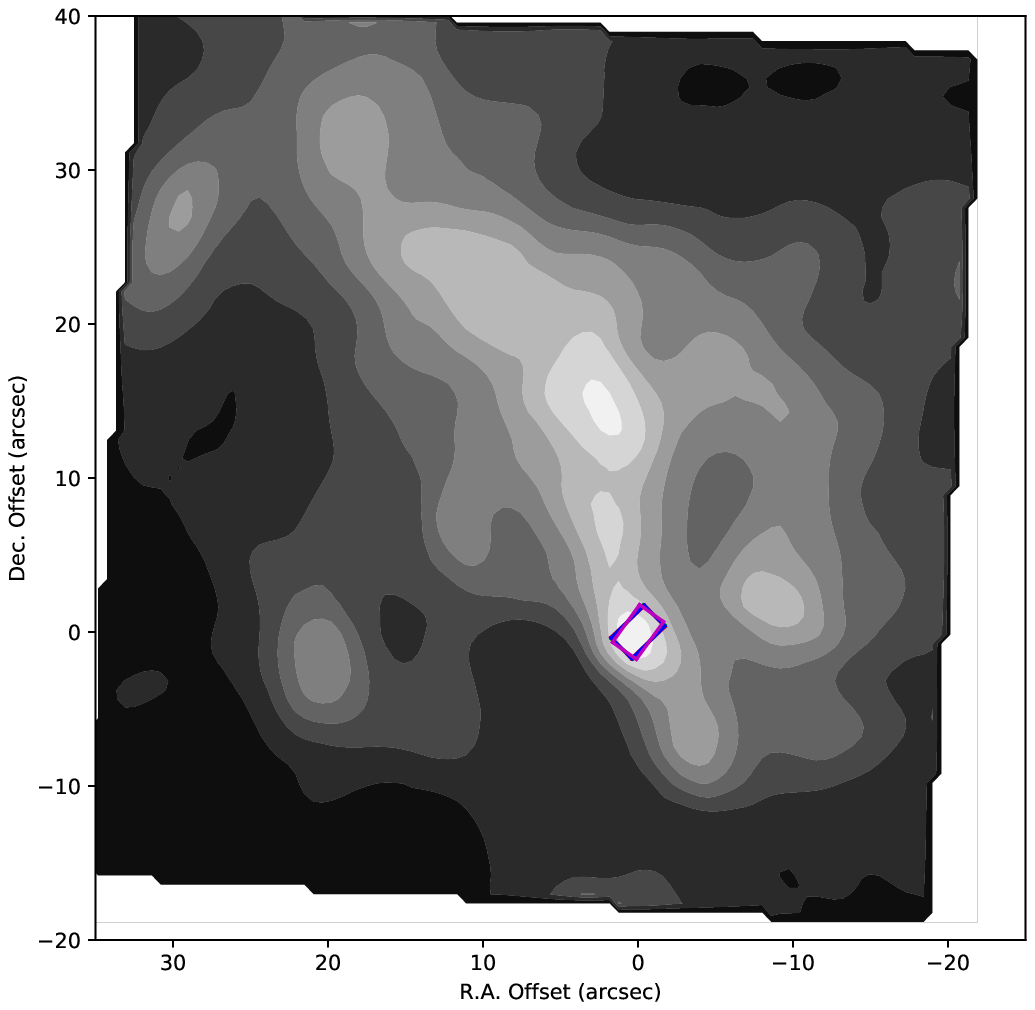}
\caption{Slit positions for IC443C.  Grayscale: map of H$_2$ S(5) intensity from 
{\it Spitzer}/IRS (Neufeld et al.\ 2007).  \re{Colored rectangles: location and orientation of  the extracted apertures for SOFIA/EXES observations of H$_2$ S(4) (red), S(5) (green),  S(6) (blue) and S(7) (magenta).}}  
\end{figure}

\begin{figure}
\includegraphics[scale=0.7,angle=0]{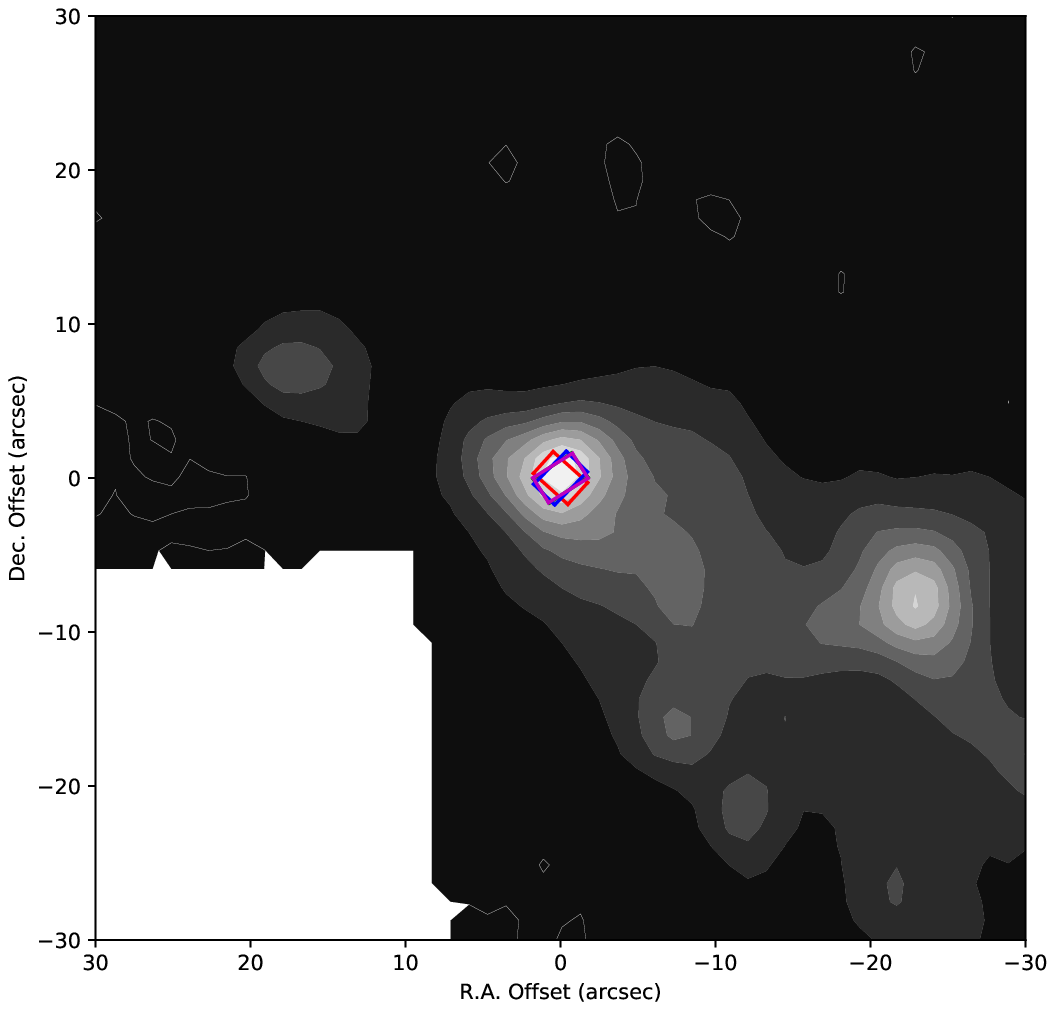}
\caption{Slit positions for NGC 2071.   Grayscale: map of H$_2$ S(5) intensity from 
{\it Spitzer}/IRS (Neufeld et al.\ 2009).  \re{Colored rectangles: location and orientation of  the extracted apertures for SOFIA/EXES observations of H$_2$ S(4) (red), S(5) (green),  S(6) (blue) and S(7) (magenta).}}
\end{figure}

\begin{figure}
\includegraphics[scale=0.7,angle=0]{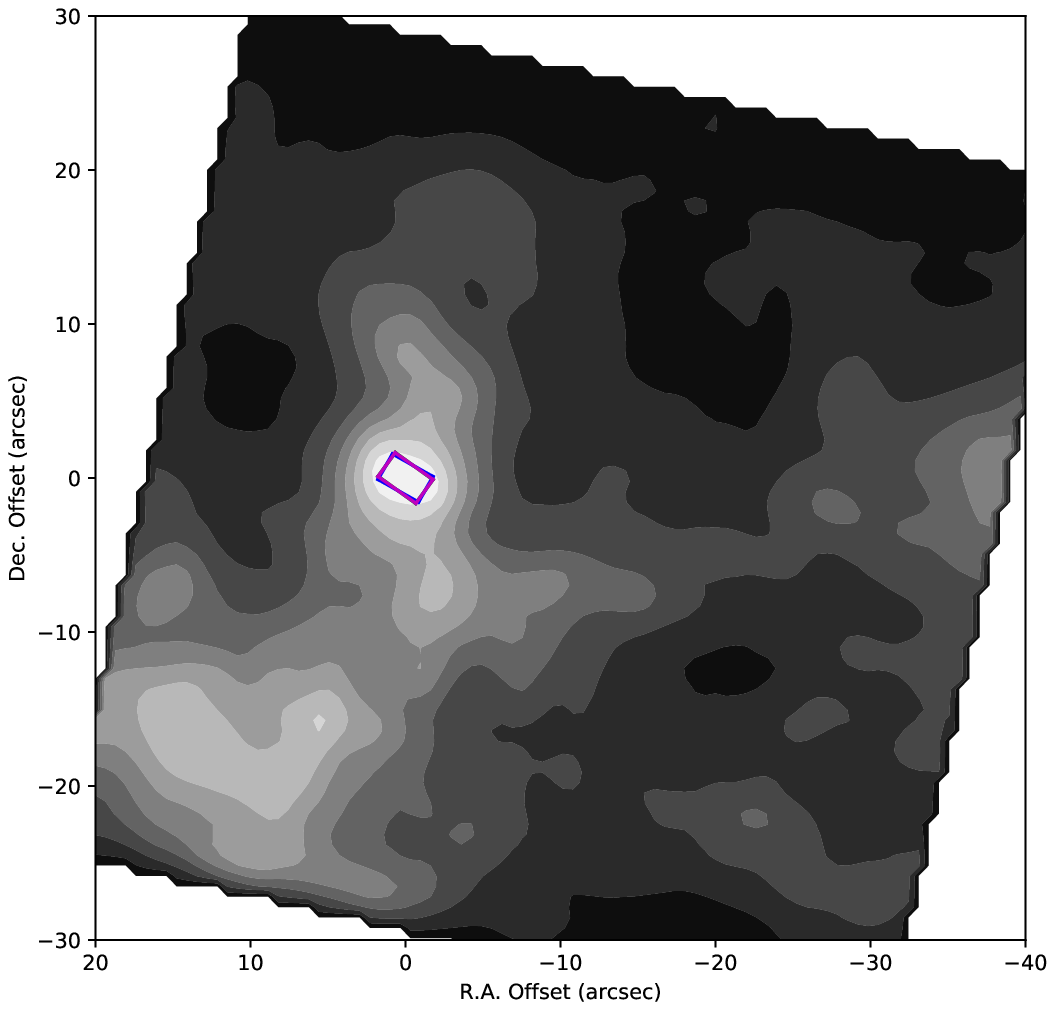}
\caption{Slit positions for 3C391.  Grayscale: map of H$_2$ S(5) intensity from 
{\it Spitzer}/IRS (Neufeld et al.\ 2007). \re{Colored rectangles: location and orientation of  the extracted apertures for SOFIA/EXES observations of S(5) (green),  S(6) (blue) and S(7) (magenta).}}
\end{figure}

\begin{table}
\scriptsize
\caption{\ma{Observational details }}
\vskip 0.1 true in
\begin{tabular}{lcccc}
\hline
Line 						& H$_2$ S(4) & H$_2$ S(5) & H$_2$ S(6) & H$_2$ S(7)		\\
							& $J=6-4$ & $J=7-5$ & $J=8-6$ & $J=9-7$ \\
Symmetry					& para & ortho & para & ortho \\
Rest wavelength ($\mu$m)	&	8.0250410 & 6.9095086 & 6.1085638 & 5.5111833		\\
Upper state energy ($E_U/k$ in K) & 3474.5 & 4586.1 & 5829.8 & 7196.7 \\
\hline
\multicolumn{5}{l}{\bf Observations of IC443C with the slit centered at 6h17m42.37s  $\bf 22^021^\prime 16.9\arcsec$ (J2000)}\\
Date (UT)	& 2022 Feb 25 & 2022 Feb 25   & 2022 Mar 1  	& 2022 Mar 7 \\
						& 2022 Mar 9 &   &  	& 		\\
Integration time$^a$ (s)& 1216		& 1024		& 1920		& 2560			\\	
& 1088	\\	
Slit position angle$^b$ & 311.9 - 309.4  & 317.0 - 315.6 & 314.7 - 309.7 & 324.3 - 315.9 \\
 & 310.6 - 307.1  \\
\hline
\multicolumn{5}{l}{\bf Observations of NGC 2071 with the slit centered at 5h47m08.21d 
$\bf 0^022^\prime 52.7\arcsec$ (J2000)}\\
Date (UT)				& 2021 Dec 2 & 2021 Dec 1  & 2022 Mar 5  	& 2022 Feb 6 \\
						& 2021 Dec 4 & 2021 Dec 2  &  	& 		\\
Integration time$^a$ (s)& 1024		& 768		& 1280		& 1280			\\	
						& 2048		& 512	\\	
Slit position angle$^b$ & 248.5 - 248.5  & 303.7 - 302.8  & 315.5 - 316.2 & 301.9 - 304.8 \\
 						& 227.9 - 227.9  & 248.5 - 248.5  \\
\hline
\multicolumn{5}{l}{\bf Observations of 3C391 with the slit centered at 18h49m22.56s $\bf -0^057^\prime 17.2\arcsec$ (J2000)}\\
Date (UT)	&  & 2022 Mar 8   & 2022 Mar 9  & 2022 Feb 24 \\
						&  &   &  	& 	2022 Mar 9	\\
Integration time$^a$ (s)& 	& 1664		& 2240		& 1920			\\	
& & & & 1280	\\	
Slit position angle$^b$ &  & 234.0 - 233.2 & 240.6 - 240.6 & 234.6 - 234.1 \\
 &  & & & 243.1 - 240.6  \\

\end{tabular}

\tablenotetext{a}{Detector integration times, on source}
\tablenotetext{b}{Degrees East of North}

\end{table} 

The observations were performed with a slit width of $1.9^{\prime\prime}$, using the EXES instrument (Richter et al.\ 2018) in High-Medium mode, a configuration that provides a spectral resolving power $\lambda/\Delta \lambda$ of 86,000 ($\Delta v = 3.5$~km~s$^{-1}$).  In Table 1, we list for each spectral line and each source the rest wavelength, slit center position, date(s) of observation, integration time, and \re{range of slit position angles} (degrees East of North) during the course of the observation.  
To enable the subtraction of telluric emission features, the telescope was nodded periodically  to a reference position devoid of H$_2$ emission. 

The data were reduced using the Redux pipeline (Clarke et al. 2015) with the fspextool software package -- a modification of the Spextool package (Cushing et al. 2004) -- which performs wavelength calibration.  The latter was obtained from observations of multiple atmospheric lines of water and 
methane within each bandpass: we conservatively estimate its accuracy as $0.3\, \rm km \, s^{-1}$. The absolute flux calibration is
accurate to $\sim \rm 25\%$ and the relative flux calibration to $\sim \rm 12.5\%$.
The atmospheric transmission is expected to exceed 90$\%$ for all observed transitions, with the exception of the S(4) line observations toward 3C391.  Because of the
unfavorable radial velocity of the source at the time of those observations, 
they were severely affected by telluric methane lines and could not be used.

\ree{All the observations of a given source were conducted within a time period of $\sim 3$ months or less.   This is much smaller than the minimum expected variability timescale
for a source at a distance $d$:  $\tau = (\theta/{\rm rad}) d /v_t = 36 (d/400{\rm \,pc}) (100\,{\rm km\,s}^{-1} / v_t) \rm\,  yr$, 
where $\theta=1.9^{\prime \prime}$ is the slit width and $v_t$ is the transverse velocity of the source.}

\section{Results}

With the exception of the S(4) line toward 3C391, all four transitions were securely detected toward each target source.  The resultant spectra are shown in Figures 4 -- 6. 
\ree{Where multiple observations were performed, the spectra were reduced individually
and then coadded with a weighting proportional to the exposure time.}
Whereas the spectra obtained by N19 toward HH7 could be adequately fitted with a single Gaussian feature, two Gaussian components are needed to fit the lines detected in the
present study.   Solid lines show the two-component fits to the observed spectra: these
use the Levenberg-Marquardt algorithm to fit two Gaussians and a first-order baseline to the data.  The velocity centroids, line widths (FWHM) and peak line intensities
for those components are presented in Table 2, along with the standard errors on each parameter.   We note that these standard errors are simply statistical in nature, and do not include uncertainties associated with flux and wavelength calibration (see Section 2 above).  Moreover, particularly for broad components, additional errors may
result from baseline ripple (i.e. departures of the baseline from the linear function
assumed here).
Vertical black lines in Figures 4 -- 6 indicate the velocity centroids for the Gaussian components, with 
horizontal bars at the bottom indicating the standard errors (i.e. 68\% confidence limits not accounting for systematic uncertainties).

\begin{table}
\scriptsize
\caption{\ma{Line fit parameters}}
\vskip 0.1 true in
\begin{tabular}{lrrrr}
\input table2.tex

\end{tabular} 
\tablenotetext{a}{Units of $10^{-3} \rm erg \, cm^{-2} \, s^{-1} \, sr^{-1} / cm^{-1}$,
\ree{where $10^{-3} \rm erg \, cm^{-2} \, s^{-1} \, sr^{-1} / cm^{-1} = \rm 3335.64\, MJy\, sr^{-1}$}}
\end{table}

\begin{figure}
\includegraphics[scale=0.6,angle=0]{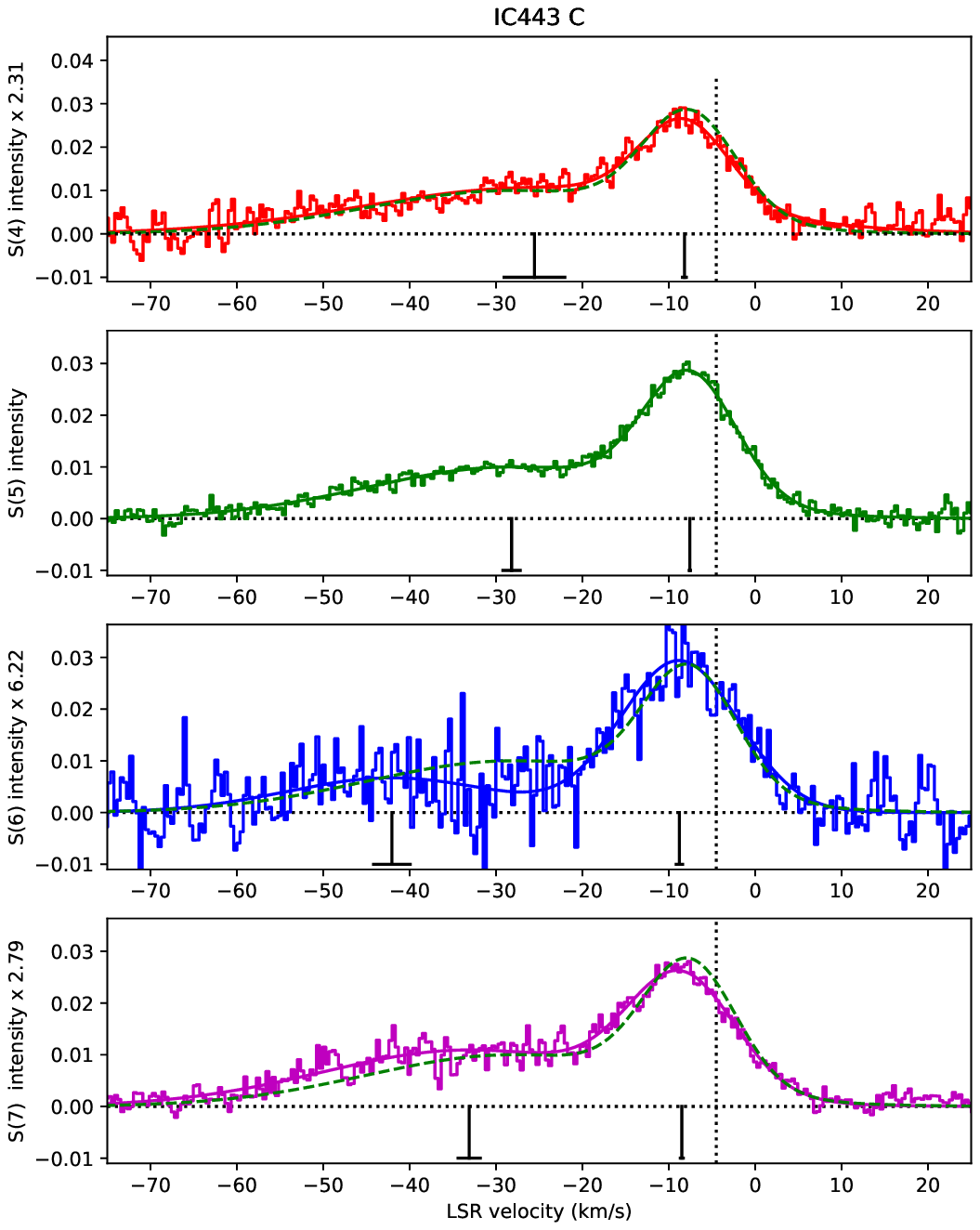}
\caption{Spectra obtained toward IC443C.  The average intensity within the $1.9 \times 3.0$ arcsec extraction region is plotted
in units of $\rm erg \, cm^{-2} \, s^{-1} \, sr^{-1} / cm^{-1}$.
Solid curves show the two-Gaussian fits to each line.  The dotted green lines show
the H$_2$ S(5) line profile, scaled and overplotted in the other panels for comparison.
\ree{Vertical black lines indicate the velocity centroids for the Gaussian components, with 
horizontal bars at the bottom indicating the standard errors}.
\bl{The LSR velocity of the source is indicated by the dotted black vertical line.}}
\end{figure}

\begin{figure}
\includegraphics[scale=0.6,angle=0]{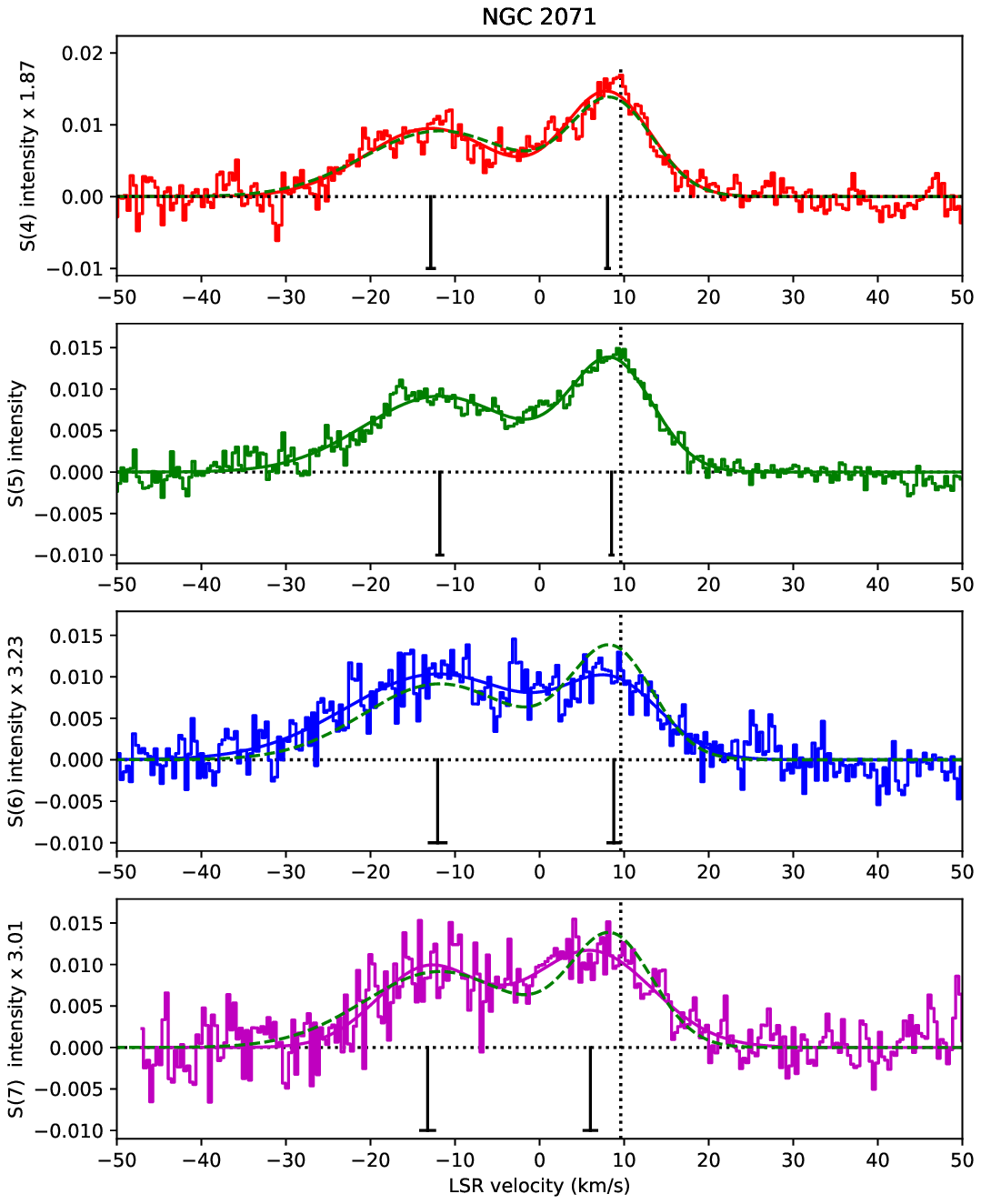}
\caption{Same as Fig.\ 4, but for NGC 2071}
\end{figure}

\begin{figure}
\includegraphics[scale=0.6,angle=0]{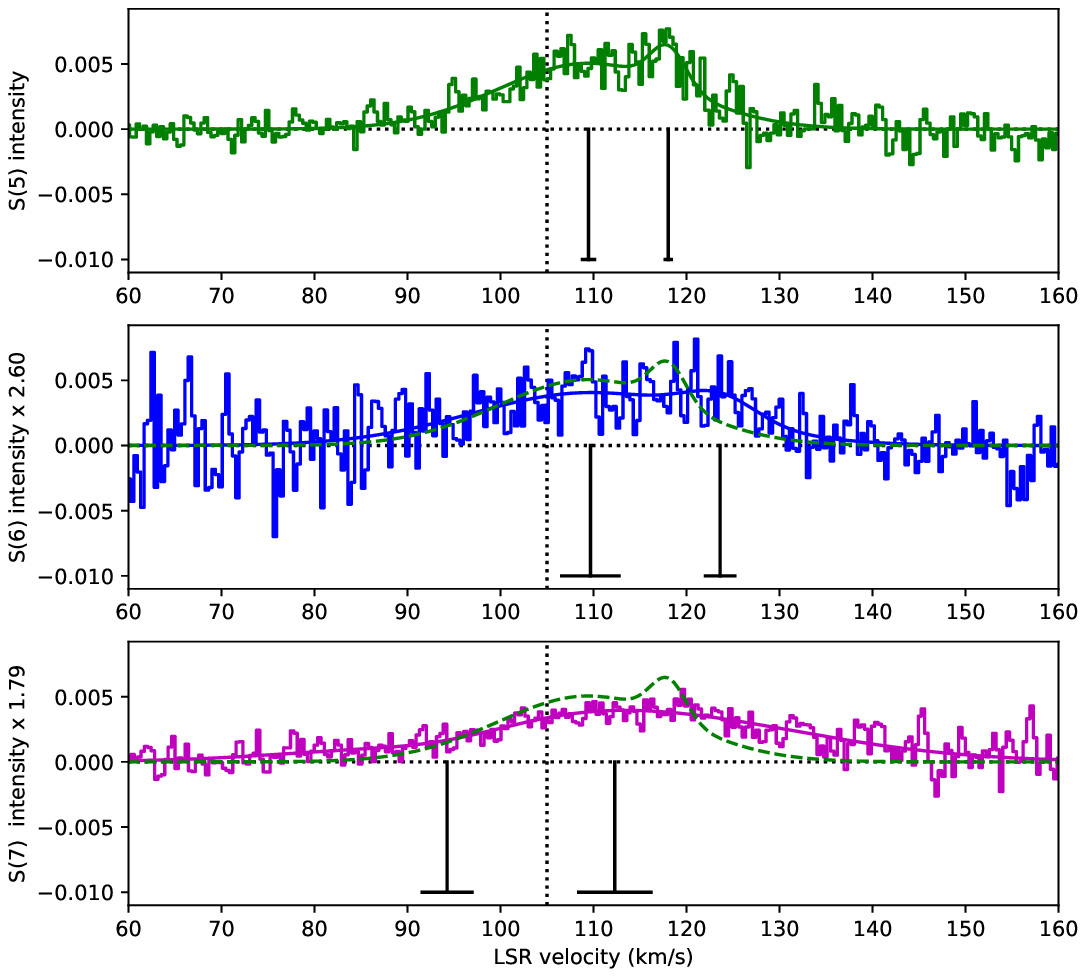}
\caption{Same as Fig.\ 4, but for 3C391.}
\end{figure}

Perhaps by coincidence, IC443C and NGC 2071 show rather similar spectra.  Each exhibits narrow component -- of best-fit FWHM $\sim 12$~km~s$^{-1}$ for the strongest, best-observed lines [i.e. S(4) and S(5)] -- and a broader component of smaller peak intensity that is blueshifted $\sim 20$~km~s$^{-1}$ relative to the narrow component.  
\re{The observed line profile for the S(5) line in IC443C has been previously modeled by Reach et al.\ (2019), who found that most of the emission could be accounted for by 
a combination of nonstationary
CJ-type shocks} \bl{consistent with the SNR expansion ram pressure (one at 60 $\rm km\,s^{-1}$ propagating in a 2000 cm$^{-3}$ clump producing the broad component, and one at 37 $\rm km\,s^{-1}$ propagating in a 6000 cm$^{-3}$ producing the narrower component, both viewed close to face-on.)}

The line emission
from 3C391 is weaker, and the observed spectrum less well-determined although also similar
to those in IC443C and NGC 2071.  The narrow components in IC443C and NGC 2071 present
the best opportunity to detect a systematic velocity shift between the ortho-H$_2$ [S(5) and S(7)] para-H$_2$ [S(4) and S(6)] lines, but none is observed reliably.  The shift between the best-fit velocity centroids for the S(4) and S(5) lines is only $0.6 \pm 0.2$ and $0.5 \pm 0.3$~km~$s^{-1}$ respectively for the narrow components in IC443C and NGC 2071.
These values are at best of marginal statistical significance -- even neglecting possible systematic uncertainties -- and are far smaller than the $3-4 \rm \, km\,s^{-1}$ shift discovered (N19) in HH7.  \re{A caveat applies to the comparison between
the S(4) and S(5) line profiles in NGC 2071, because the slit position angles are significantly different for the S(4) line observation and the second S(5) observation (accounting for $40\%$ of the S(5) observing time).  Any systematic difference this might lead to is mitigated by the fact that the $1.9 \times 3.0$ arcsec extraction region is only modestly elongated with an aspect ratio of 1.58.}

\section{Discussion}

Because the quadrupole H$_2$ rotational lines are always optically thin, the intensity of 
the S(X-2) line is linearly proportional to the column density in the upper $J=$~X state. 
Following N19, we may conveniently define excitation temperatures, $T_{XY}$, that 
characterize the column densities ratios measured between pairs of states $J=$~X and $J=$~Y of the same spin symmetry (i.e. with X and Y both odd or both even.)  And given a triad of measured
column densities for three consecutive $J$-states, X, Y, and Z, we may define the ortho-to-para, ${\rm OPR}_{XYZ}$, that is needed to account for the $J=$~Y level population based on a linear interpolation of the rotational diagram between $J=$~X and $J=$~Z.  The results for the four parameters $T_{86}$, $T_{97}$, ${\rm OPR}_{678}$, and 
${\rm OPR}_{789}$ are shown in Figures 7 -- 9 (crosses) in LSR velocity bins of width 2~km~s$^{-1}$, along with a normalized plot
showing the rebinned spectra.  \bl{For each parameter, the average values are
indicated both with text and a dashed colored line.  The individual points have 
error bars that reflect the noise in the spectrum but not the flux calibration uncertainties; here, the r.m.s.\ noise was propagated with a Monte Carlo method.
The average values indicated with text, by contrast, {\it do} include the systematic
uncertainties associated with flux calibration (see Section 2 above).  Dashed black lines indicate the
values (both 3.00) that ${\rm OPR}_{678}$ and ${\rm OPR}_{789}$ would acquire in thermal
equilibrium at the inferred gas temperatures $T_{86}$ and $T_{97}$}

Whereas the results presented by N19 for HH7 showed a steady increase
of ${\rm OPR}_{678}$ from a value of $\sim 0.5$ at the red edge to $\sim 2$ at the blue edge, in the sources studied in the present paper there is no strong evidence for a systematic variation of the OPR with LSR velocity.  This negative result, of course, is
equivalent to the finding (noted in Section 3) that any velocity shifts between
ortho- and para-H$_2$ are small.
There are, however, systematic differences between $T_{86}$ and $T_{97}$ and between ${\rm OPR}_{678}$ and 
${\rm OPR}_{789}$ (with the sense that \bl{$T_{86} < T_{97}$ and ${\rm OPR}_{678} < {\rm OPR}_{789}$)}.  These are both a natural consequence of the positively-curved rotational diagram that can result when there is a mixture of gas temperatures along the sightline.
\bl{Because the definition of ${\rm OPR}_{XYZ}$ (see above) is based on a {\it linear } interpolation of the rotational diagram between the $J=\rm X$ and $J= \rm Z$ states,
the effect of positive curvature will tend to overestimate the OPR when Y is even 
(i.e.\ when $J=Y$ is a para-state) and underestimate the OPR when Y is odd. 
We therefore expect the actual OPR in the hot gas to be bracketed by 
${\rm OPR}_{678}$ and ${\rm OPR}_{789}$}.

\begin{figure}
\includegraphics[scale=0.7,angle=0]{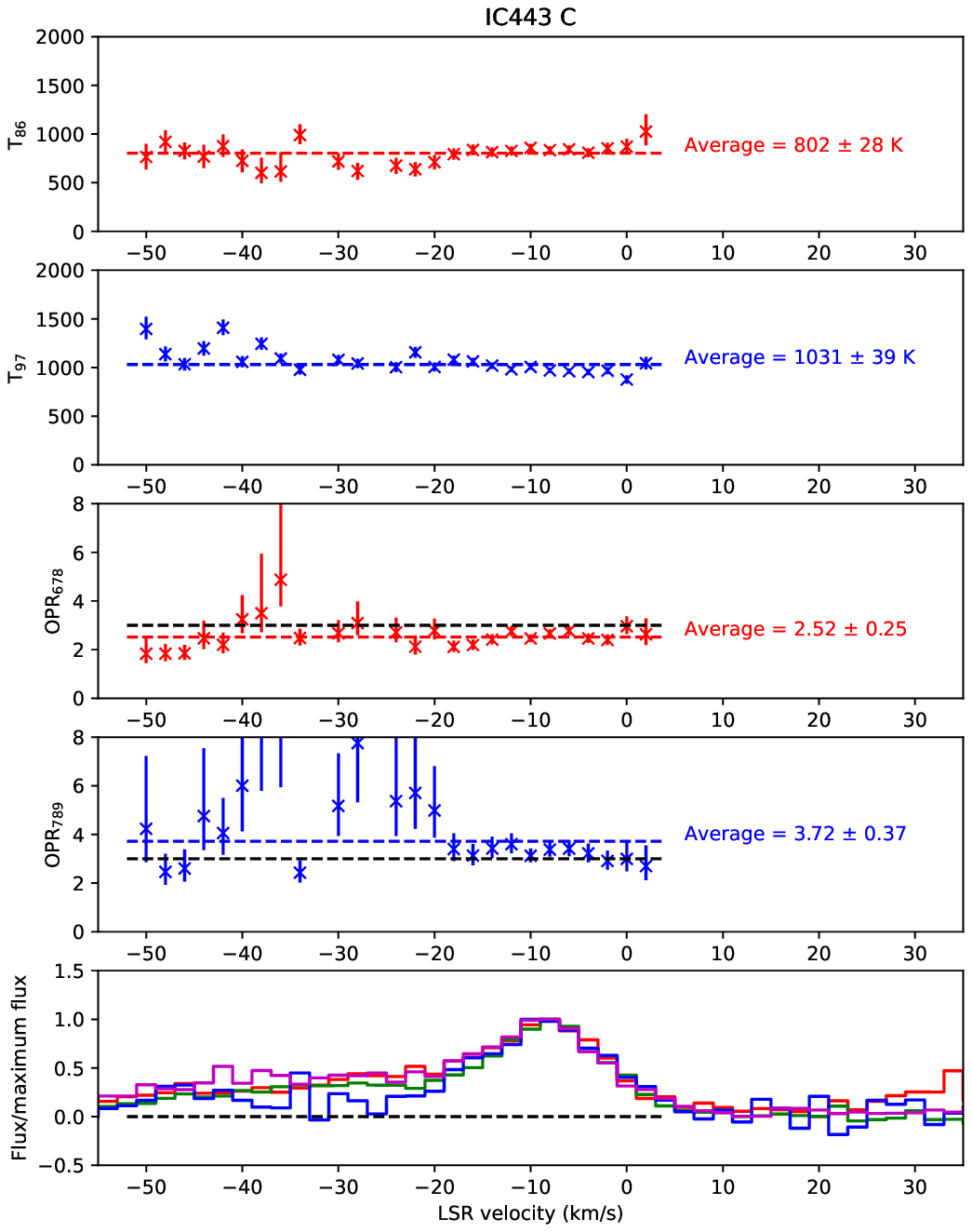}
\caption{Results obtained for IC443C.  Top two panels: $T_{XY}$, the temperature derived
from the relative populations in the H$_2$ $J=$~X and Y states.  Next two panels: 
OPR$_{XYZ}$, the ortho-to-para ratio derived
from the relative populations in the H$_2$ $J=$~X, Y and Z states.  Bottom panels: scaled spectra, rebinned to 2~km~s$^{-1}$ resolution, with the same color-coding as in  
Figures 4 -- 6.}
\end{figure}

\begin{figure}
\includegraphics[scale=0.7,angle=0]{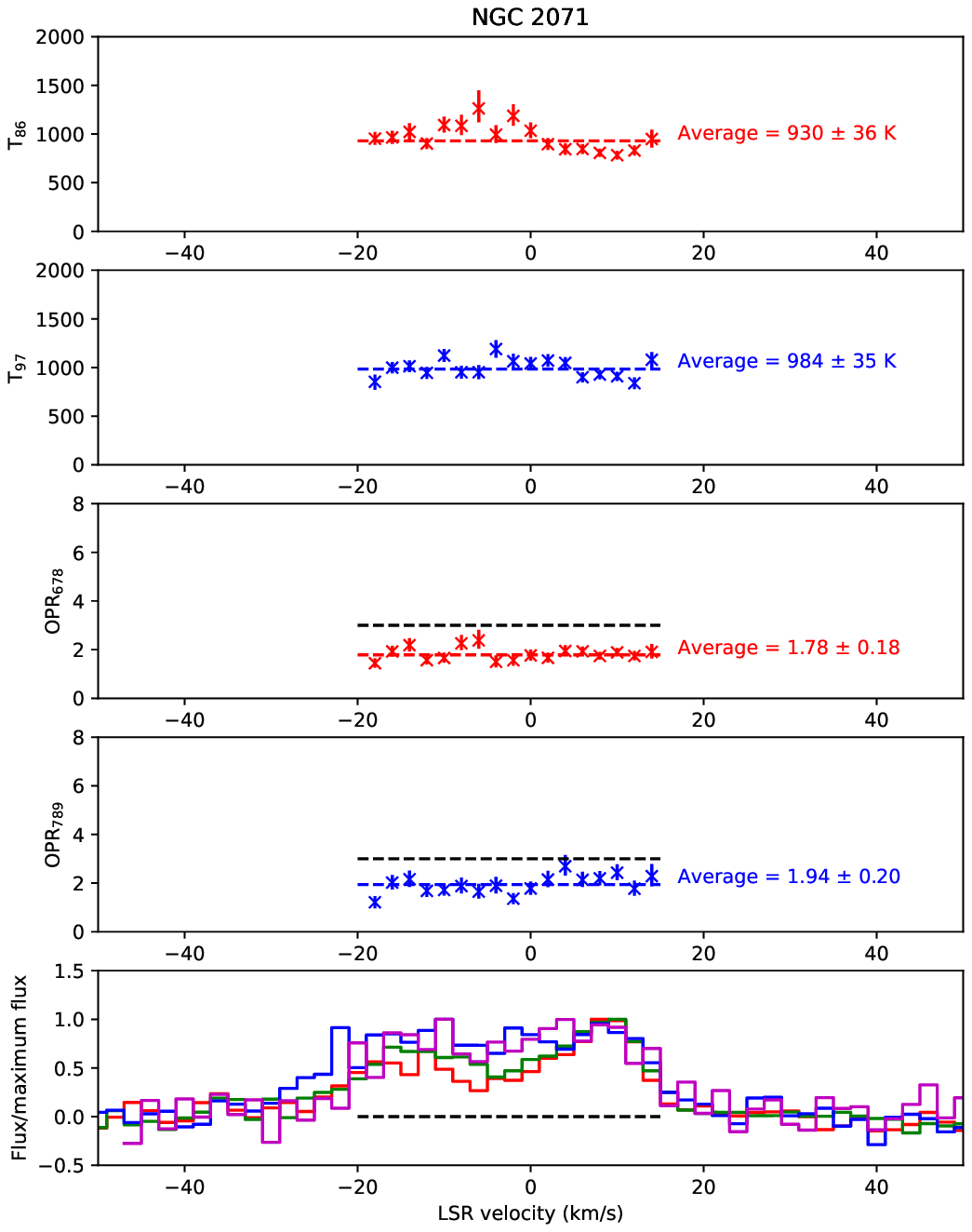}
\caption{Results obtained for NGC 2071.  Top two panels: $T_{XY}$, the temperature derived
from the relative populations in the H$_2$ $J=$~X and Y states.  Next two panels: 
OPR$_{XYZ}$, the ortho-to-para ratio derived
from the relative populations in the H$_2$ $J=$~X, Y and Z states.  Bottom panels: scaled spectra, rebinned to 2~km~s$^{-1}$ resolution, with the same color-coding as in  
Figures 4 -- 6.}
\end{figure}

\begin{figure}
\includegraphics[scale=0.7,angle=0]{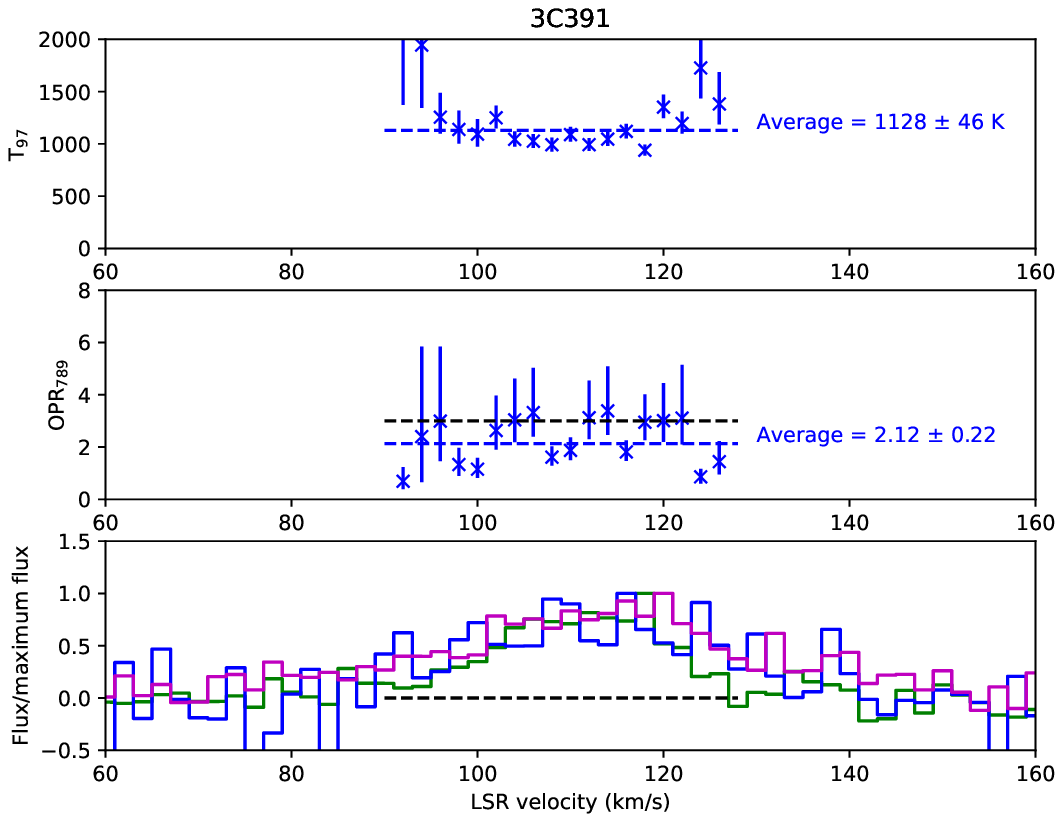}
\caption{Results obtained for 3C 391.  Top panel: $T_{97}$, the temperature derived
from the relative populations in the H$_2$ $J=$~7 and 9 states.  Middle panel: 
OPR$_{789}$, the ortho-to-para ratio derived
from the relative populations in the H$_2$ $J=$~7, 8 and 9 states.  Bottom panels: scaled spectra, rebinned to 2~km~s$^{-1}$ resolution, with the same color-coding as in  
Figures 4 -- 6.}
\end{figure}

\re{Several factors may contribute to the absence of detectable ortho-para velocity shifts
in these sources.  In the case of IC443C, which was selected as a control against which the other sources were to be compared with EXES, previous {\it Spitzer/IRS} observations  of the H$_2$ S(0) - S(7) lines (Neufeld et al. 2007) had revealed no significant departures from an OPR of 3.  In this source, the shock is likely 
propagating in material that {\it already} possesses an OPR close to the high-temperature equilibrium value of 3.  Ortho-to-para conversion occurs only slowly within the cold preshock gas; thus, if the material suffers repeated heating events in shocks that drive the OPR towards its high-temperature equilibrium value, then there may be insufficient time to reset the OPR to a low temperature value.}  \bl{This possibility is supported
by HI (Lee et al. 2008) and optical (DSS, York et al. 2000) observations of IC443 that show filamentary structures lying further from the center of the SNR than clump C, suggesting that the molecular shell to which IC443C belongs has already been processed by shocks.}  

\re{For NGC 2071 and 3C391, by contrast, the lowest $J$ states observed with Spitzer imply the presence of a warm gas component with an OPR significantly below the LTE value.  The H$_2$ 
line fluxes in these sources (Giannini et al. 2011; Neufeld et al.\ 2009) are well accounted for by a two component model that invokes a warm (250 - 350 K) gas component
responsible for the lower-lying transitions and a hot (1000 - 1100 K) component
responsible for the higher-lying transitions: such models
yield best estimates of 0.52 and 0.65 for the OPR in the warm components in NGC 2071 and 3C391, respectively.  Although not as extreme as the value of 0.21 obtained for the
warm component in HH7 (Neufeld et al.\ 2007), these values imply the presence of
preshock gas with a low-temperature OPR.}  \bl{As was widely observed in {\it Spitzer} observations of shocked regions (e.g.\ Neufeld et al.\ 2006, 2007, 2009), the hot gas
component in these sources, primarily traced by the H$_2$ S(4) -- S(7) lines, exhibits a
larger OPR than the warm component.  This behavior likely reflects the fact (Neufeld et al.\ 2006) that para-to-ortho conversion proceeds primarily through
reactive collisions with atomic hydrogen that possess
a significant activation energy barrier ($E_A/k \sim 3900$~K).
In this picture, multiple shocks with different velocities
are present within the beam, with the lower velocity
shocks leading to lower temperature gas within which the timescale for 
complete para-to-ortho conversion exceeds the period for which the gas remains warm.}

Because NGC 2071 and 3C391 both possess a warm gas component with an OPR far
below the equilibrium value, we had expected them to be excellent targets
for the detection of ortho-para velocity shifts like those observed previously in HH7.
The absence of such shifts in the SOFIA data reported here may reflect the
presence of a different
geometry than that in HH7.  The shocked gas in NGC 2071 and 3C391 lacks the well-defined bow shock morphology apparent in HH7, raising the possibility that multiple shocks with different propagation directions are present within the slit, or along the line-of-sight,
\bl{and contribute to the S(4) -- S(7) line emissions}.
This possibility also becomes more likely because the projected apertures are considerably larger than in HH7, owing to the
greater source distances (estimated as 0.39 and 9 kpc respectively 
for NGC 2071 and 3C391, as compared to 0.22 kpc for HH7); \ree{observations at
higher spatial resolution (but poorer spectral resolution) with JWST would be
complementary here and could reveal structures on scales smaller than the EXES slit size}.

\ree{Although detailed calculations will be presented in a future study (Reach et al. in preparation), any model in which the broad observed lines are attributed to velocity gradients within a single shock is expected to show an ortho-para shift if the initial OPR is low enough, regardless of whether it invokes C- or CJ-shocks and whether the model is stationary or non-stationary.  The one exception here might be a hot and partly dissociative shock in which the high temperature resets the OPR before the bulk
of the H$_2$ emission occurs.}

\begin{acknowledgements}

Based on observations made with the NASA/DLR Stratospheric Observatory for Infrared Astronomy (SOFIA). SOFIA was jointly operated by the Universities Space Research Association, Inc. (USRA), under NASA contract NAS2-97001, and the Deutsches SOFIA Institut (DSI) under DLR contract 50 OK 0901 to the University of Stuttgart. D.A.N gratefully acknowledges the support of a SOFIA/USRA grant, SOF09-0021; W.T.R. that of SOFIA/USRA grant 07-0007;
and S.C., A.G., and P.L.\ that of the Programme National ``Physique et
Chimie du Milieu Interstellaire" (PCMI) of CNRS/INSU with INC/INP co-funded by CEA and CNES.  We are grateful to the EXES PI, Matt Richter, and the EXES team for their support of the observations presented here.

\end{acknowledgements}

\end{document}

%% file: table2.tex
{\bf Results for IC443 C} \cr
\hline
Line & S(4) \phantom{00} & S(5) \phantom{00} & S(6)\phantom{00} & S(7) \phantom{00}\cr
$\lambda_{\rm rest} (\mu$m) & 8.02504108 & 6.90950858 & 6.10856384 & 5.511183259 \cr
\hline
$v_1$ (km/s) & $-8.2 \pm 0.2$ & $-7.6 \pm 0.1$ & $-8.8 \pm 0.4$ & $-8.5 \pm 0.2$ \cr 
$\Delta v_1$ (km/s FWHM) & $12.1 \pm 0.9$ & $12.6 \pm 0.3$ & $16.1 \pm 1.0$ & $14.6 \pm 0.6$ \cr 
$I_1$ (Note a) & $8.3 \pm 0.6$ & $24.0 \pm 0.6$ & $4.7 \pm 0.2$ & $8.0 \pm 0.3$ \cr 
$v_2$ (km/s) & $-25.6 \pm 3.5$ & $-28.2 \pm 1.0$ & $-42.1 \pm 2.1$ & $-33.1 \pm 1.3$ \cr 
$\Delta v_2$ (km/s FWHM) & $46.3 \pm 8.4$ & $39.4 \pm 2.5$ & $28.7 \pm 6.8$ & $40.2 \pm 3.9$ \cr 
$I_2$ (Note a) & $4.6 \pm 0.7$ & $10.0 \pm 0.3$ & $1.1 \pm 0.2$ & $3.9 \pm 0.2$ \cr 
\hline
\cr
{\bf Results for NGC 2071} \cr
\hline
Line & S(4) \phantom{00} & S(5) \phantom{00} & S(6)\phantom{00} & S(7) \phantom{00}\cr
$\lambda_{\rm rest} (\mu$m) & 8.02504108 & 6.90950858 & 6.10856384 & 5.511183259 \cr
\hline
$v_1$ (km/s) & $8.0 \pm 0.2$ & $8.5 \pm 0.2$ & $8.8 \pm 0.7$ & $6.0 \pm 0.7$ \cr 
$\Delta v_1$ (km/s FWHM) & $12.0 \pm 0.6$ & $11.5 \pm 0.4$ & $13.5 \pm 1.5$ & $17.3 \pm 1.8$ \cr 
$I_1$ (Note a) & $7.7 \pm 0.2$ & $13.1 \pm 0.3$ & $2.6 \pm 0.3$ & $3.9 \pm 0.2$ \cr 
$v_2$ (km/s) & $-12.9 \pm 0.4$ & $-11.8 \pm 0.4$ & $-12.1 \pm 1.0$ & $-13.2 \pm 0.8$ \cr 
$\Delta v_2$ (km/s FWHM) & $18.2 \pm 1.2$ & $21.2 \pm 1.0$ & $25.4 \pm 2.6$ & $14.2 \pm 1.8$ \cr 
$I_2$ (Note a) & $5.1 \pm 0.2$ & $9.2 \pm 0.2$ & $3.2 \pm 0.1$ & $3.2 \pm 0.2$ \cr 
\hline
\cr
{\bf Results for 3C391} \cr
\hline
Line & S(4) \phantom{00} & S(5) \phantom{00} & S(6)\phantom{00} & S(7) \phantom{00}\cr
$\lambda_{\rm rest} (\mu$m) & 8.02504108 & 6.90950858 & 6.10856384 & 5.511183259 \cr
\hline
$v_1$ (km/s) & N/A & $118.0 \pm 0.3$ & $123.6 \pm 1.6$ & $112.3 \pm 3.9$ \cr 
$\Delta v_1$ (km/s FWHM) & N/A & $4.5 \pm 0.9$ & $9.2 \pm 5.7$ & $44.5 \pm 7.6$ \cr 
$I_1$ (Note a) &  N/A  & $3.0 \pm 0.5$ & $0.7 \pm 0.4$ & $2.2 \pm 0.3$ \cr 
$v_2$ (km/s) &  N/A & $109.5 \pm 0.6$ & $109.7 \pm 3.1$ & $94.3 \pm 2.7$ \cr 
$\Delta v_2$ (km/s FWHM) & N/A & $23.0 \pm 1.3$ & $30.0 \pm 6.1$ & $18.1 \pm 11.7$ \cr 
$I_2$ (Note a) &  N/A & $5.1 \pm 0.2$ & $1.6 \pm 0.2$ & $-0.5 \pm 0.5$ \cr 
\hline